Visualizing the context of citations

referencing papers published by Eugene Garfield:

A new type of keyword co-occurrence analysis

Lutz Bornmann*, Robin Haunschild**, and Sven E. Hug***

*Corresponding author:

Division for Science and Innovation Studies

Administrative Headquarters of the Max Planck Society

Hofgartenstr. 8,

80539 Munich, Germany.

Email: bornmann@gv.mpg.de

** Max Planck Institute for Solid State Research,

Heisenbergstr. 1,

70569 Stuttgart, Germany.

Email: R.Haunschild@fkf.mpg.de

***Evaluation Office

University of Zurich

Mühlegasse 21

8001 Zurich, Switzerland

Email: sven.hug@evaluation.uzh.ch

**Abstract**

During Eugene Garfield's (EG's) lengthy career as information scientist, he published about 1,500 papers. In this study, we use the impressive oeuvre of EG to introduce a new type of bibliometric networks: keyword co-occurrences networks based on the context of citations, which are referenced in a certain paper set (here: the papers published by EG). The citation context is defined by the words which are located around a specific citation. We retrieved the citation context from Microsoft Academic. To interpret and compare the results of the new network type, we generated two further networks: co-occurrence networks which are based on title and abstract keywords from (1) EG's papers and (2) the papers citing EG's publications. The comparison of the three networks suggests that papers of EG and citation contexts of papers citing EG are semantically more closely related to each other than to titles and abstracts of papers citing EG. This result accords with the use of citations in research evaluation that is based on the premise that citations reflect the cognitive influence of the cited on the citing publication.

# 1 Introduction

Assessments based on bibliometric data is the cornerstone of modern research evaluation processes (Bornmann, 2017). Research evaluation without bibliometrics seems no longer imaginable today. The foundation for this importance of bibliometrics was laid by Eugene Garfield (EG). According to Wouters (2017), EG "enabled an entire field: scientometrics, the quantitative study of science and technology". EG conceptualized a scientific citation index similarly designed as Shepard's Citations – a system which tracked how US court cases cited former ones – and published this concept in *Science* (Garfield, 1955). His invention of the index "revolutionized the world of scientific information and made him one of the most visionary figures in information science and scientometrics" (van Raan & Wouters, 2017). He founded the Institute for Scientific Information (ISI) in the 1960s (now Clarivate Analytics, see clarivate.com) which provided later the Science Citation Index, the Social Sciences Citation Index, and the Arts and Humanities Citation Index – the predecessors of the Web of Science (WoS). With algorithmic historiography, Garfield, Sher, and Torpie (1964) developed – based on the concept of citations as recursive operations – one of the first bibliometric network types which finally led to the network program HistCite™ (Leydesdorff, 2010).

During EG's lengthy career as information scientists, he published more than 1,500 publications (see www.researcherid.com/rid/A-1009-2008). In this study, we use the impressive oeuvre of EG to introduce a new type of bibliometric networks: keyword co-occurrence networks based on the context of citations, which are referenced in a certain paper set (here: the papers published by EG). The citation context is defined by the words which are located around a specific citation. Most of the networks published hitherto are citation, co-citation, bibliographic coupling, co-authorship networks, or keyword co-occurrence (Guevara, Hartmann, Aristarán, Mendoza, & Hidalgo, 2016). As a rule, previous keyword co-occurrence networks are based on extracted words from the title and abstract of a publication or from author-generated keyword lists (van Eck & Waltman, 2014). To the best of our knowledge, the citation context was never used in earlier studies as data source for generating keyword co-occurrence networks.

## 2       Literature overview and research questions

Bornmann and Daniel (2008) published an extensive review of studies investigating the context of citations (see also Zhang, Ding, & Milojević, 2013). Most of these studies have been published more than twenty years ago. The studies used the citation context to categorize citations (e.g., the citer suggests that the cited paper is erroneous in part and offers a correction). However, Bertin, Atanassova, Sugimoto, and Lariviere (2016) note that "over the last few year[s], we are witnessing an increasing body of literature on the citation context of scientific papers" (p. 1418). For example, Small, Tseng, and Patek (2017) analyzed the context of citations by searching for words that labeled referenced items as discoveries. The citing sentences which contained "discovery words" have been called "discovery citance". Hu, Chen, and Liu (2015) investigated the context of recurring citations in papers published by the *Journal of Informetrics*. They found that "for a specific reference, its first-time citation is usually not as intentional as the succeeding citations" (p. 221). Jha, Jbara, Qazvinian, and Radev (2016) used comprehensive datasets for their citation context analyses. Their results show – among other things – that the citation context can contain both evaluative and informative signals.

One of the main challenges of citation context studies is the great effort of producing the data: The text around the citations must be delimited, extracted, and categorized. In the past, this had to be done manually or semi-manually and required a lot of effort. As a consequence, not many studies have been published and most studies were small-scale. In the age of big scholarly data, producing suitable data is becoming easier. For example, there exist large text corpora with extensive XML mark-up that facilitate the analysis of citation contexts (see Bertin, et al., 2016). According to Halevi and Moed (2013) "nowadays electronic publishing enables the creation of very large text files that can be analyzed in an automatic way. Computer scientists, linguistic experts, and information scientists have begun processing the full texts of research articles in large numbers of journals" (p. 1904).

Since a few months, citation contexts can be directly downloaded from a new citation database, Microsoft Academic (MA, see Hug & Brändle, 2017; Hug, Ochsner, & Brändle, 2017). To the best of our knowledge, MA is the first large-scale scholarly database that allows downloading citation contexts which are already segmented. In this study, we use citation context data from MA to produce keyword co-occurrence networks which are based on extracted keywords from the citation context. To interpret the results of the new network type, we generated two further networks for comparison: co-occurrence networks which are based on title and abstract keywords from the (1) cited and (2) citing publications. We are interested in similarities and differences between the networks. We assume that the citation context network is similar to the network which is based on title and abstract keywords from the cited (but not citing) publications. The use of citations in research evaluation is based on the premise that citations reflect the cognitive influence from the cited to the citing publication. Thus, the citations in the citing documents should be similarly contextualized as the cited papers are described in title and abstract.

## 3 Methods

### 3.1 Microsoft Academics (MA)

MA was publicly released in February 2016 (Microsoft, 2017a) and models "the real-life academic communication activities as a heterogeneous graph" (Sinha et al., 2015, p. 244). It gets the majority of its data from web pages indexed by Bing (Sinha, et al., 2015) and is updated on a weekly basis (Microsoft, 2017b). The database is independent of its predecessor,

Microsoft Academic Search, which has been completely decommissioned towards the end of 2016. MA data can either be accessed by using the Microsoft Academic search engine[1] or by employing the Academic Knowledge API (AK API)[2]. In contrast to Google Scholar, data suitable for bibliometric analyses can be retrieved with relative ease from MA (see Hug, et al., 2017). The database is still under development and evolving quickly (see Hug & Brändle, 2017). For example, a social network for academics has been added recently. Furthermore, the database has expanded rapidly from 83 million records in 2015, to 140 million in 2016, and to 168 million in early 2017 (Hug & Brändle, 2017).

**3.2   Dataset used**

All MA data were collected via the AK API using the *Evaluate* REST endpoint (evaluates a query expression and returns AK entity results). First, we retrieved all publications by EG and their MA ID. Only 327 of EG's 1,558 publications were found in MA. Second, we retrieved all publications citing EG's papers. Citations contexts were available for 59 papers. Third, for the 59 papers and their citing papers (n=343 papers), we retrieved the title and abstract information from the WoS by searching for the Digital Object Identifier (DOI). These two sets of publications show an overlap of two papers. Finally, we extracted the citation context (428 citation contexts due to multiple citations in some citing papers) of the citing papers.

**3.3   Keyword co-occurrence analysis**

We produced three co-occurrences networks using MA data and the VOSviewer software (www.vosviewer.com; van Eck & Waltman, 2010): (1) title and abstract network based on EG's papers, (2) title and abstract network based on papers citing EG's papers, and (3) network based on citation contexts around citations of EG's papers.

---

[1] http://academic.research.microsoft.com
[2] https://www.microsoft.com/cognitive-services/en-us/academic-knowledge-api

To extract keywords from the titles, abstracts and citation contexts, the text mining function of the VOSviewer (van Eck & Waltman, 2011) was used. This function creates a co-occurrence network of keywords (adjectives and nouns) and displays it on a two-dimensional map. Two keywords are said to co-occur if they both occur in the same title/abstract or citation context. The distance between two keywords (two nodes) is approximately inversely proportional to the similarity (relatedness in terms co-occurrence) of the keywords. Hence, keywords with a higher rate of co-occurrence tend to be found closer to each other. The VOSviewer provides a clustering function, which assigns keywords to clusters based on their co-occurrence (see van Eck & Waltman, 2017; Waltman & van Eck, 2013; Waltman, van Eck, & Noyons, 2010).

To generate each of the three maps in this study, identical settings in the VOSviewer were applied: we used binary counting, a keyword had to occur at least four times, and we included the 60% most relevant keywords in the network. For each map, the number of clusters was determined based on interpretability reasons. Keywords not relevant to our analysis were excluded manually. Words that structure abstracts (e.g. 'practical implications', 'originality value') and names of cited authors in citation contexts (e.g. 'Moed et al.') were removed.

## 4 Results

In the first step of the analysis, we generated a network based on EG's papers for which a citation context was available (n=59) to get an impression of his papers (included in this study). From the titles and abstracts of these 59 papers published by EG, 607 keywords were extracted of which 27 occurred four or more times. Based on the criteria mentioned in section 3.3, 15 keywords were included in the map. The two clusters 'Journal Impact Factor' (JIF) and 'historiographs' (comprising the HistCite software package as well as the Science

Citation Index) emerged (see Figure 1). These entities form an important part of EG's legacy (see e.g. Small, 2017).

However, comparing Figure 1 with the 23 broad topics extracted by Sen (2014) from EG's publications reveals that the topics in Figure 1 are not representative for EG's extensive oeuvre. Likely, this is due to the very low number of EG's publications covered by MA, which form the basis of our analysis. Although 327 of EG's 1,558 publications were found in MA, citation contexts were available for only 59 publications. However, as our aim is not to give a complete description of EG's work, but to explore a new approach to citation contexts, the scarcity of available data is not a problem for our analysis. All three networks which we generated in this study are only related to the restricted publication set. Thus, we expect that also the citing papers of the 59 publications and the corresponding citation contexts reflect only a part of EG's influence or impact.

Figure 1. Co-occurrence of keywords in titles and abstracts of publications by Eugene Garfield

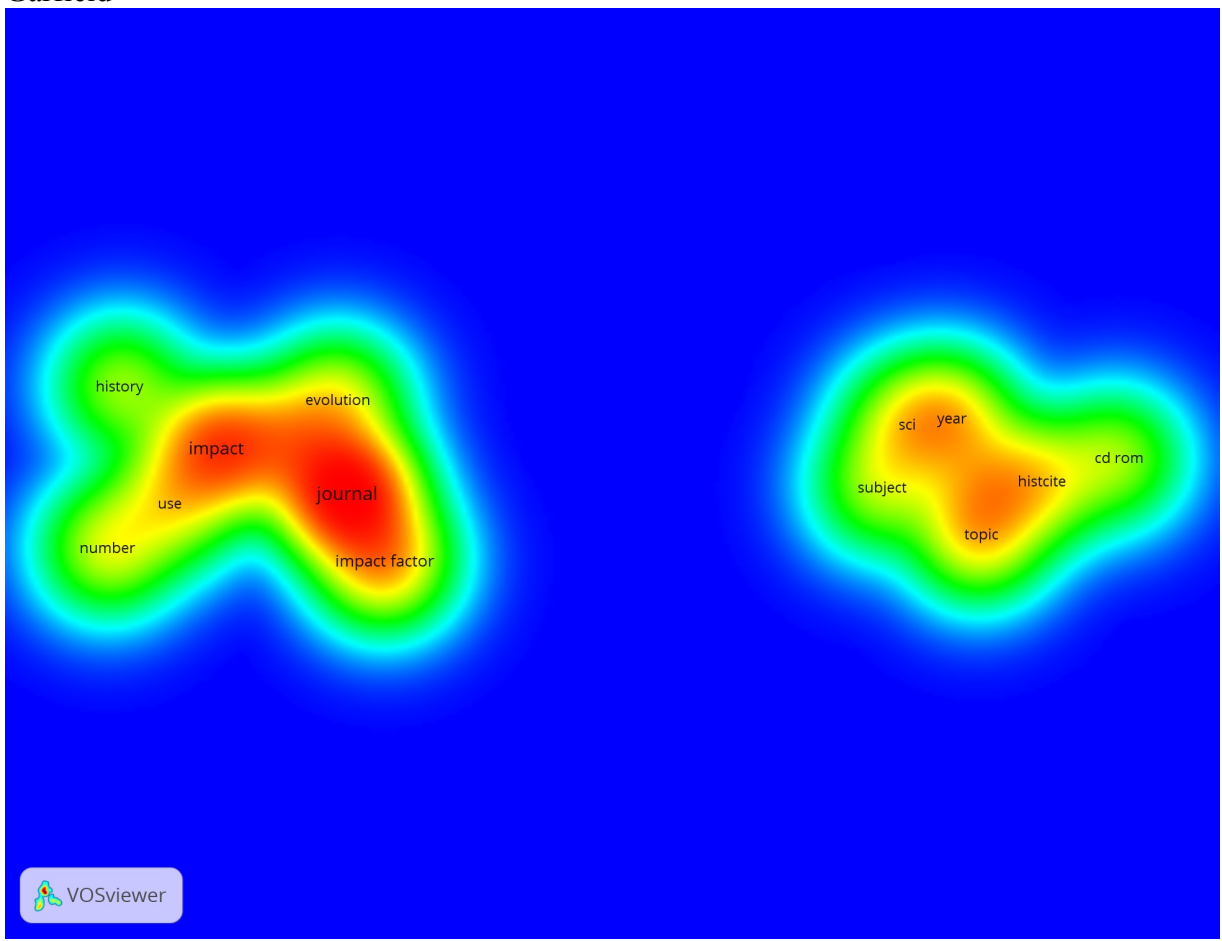

In the second step of the analysis, we produced the co-occurrence network which refers to the citing papers. From the titles and abstracts of papers citing EG's 59 papers, 6,878 keywords were extracted of which 512 occurred four or more times. After applying the inclusion criteria (see section 3.3), 297 keywords were used to build the map. These are significantly more keywords than from titles and abstracts of cited papers. Correspondingly, 18 clusters were identified in contrast to only two clusters in the cited paper map (see Figure 1).

Figure 2. Co-occurrence of keywords in titles and abstracts of papers referencing Eugene Garfield

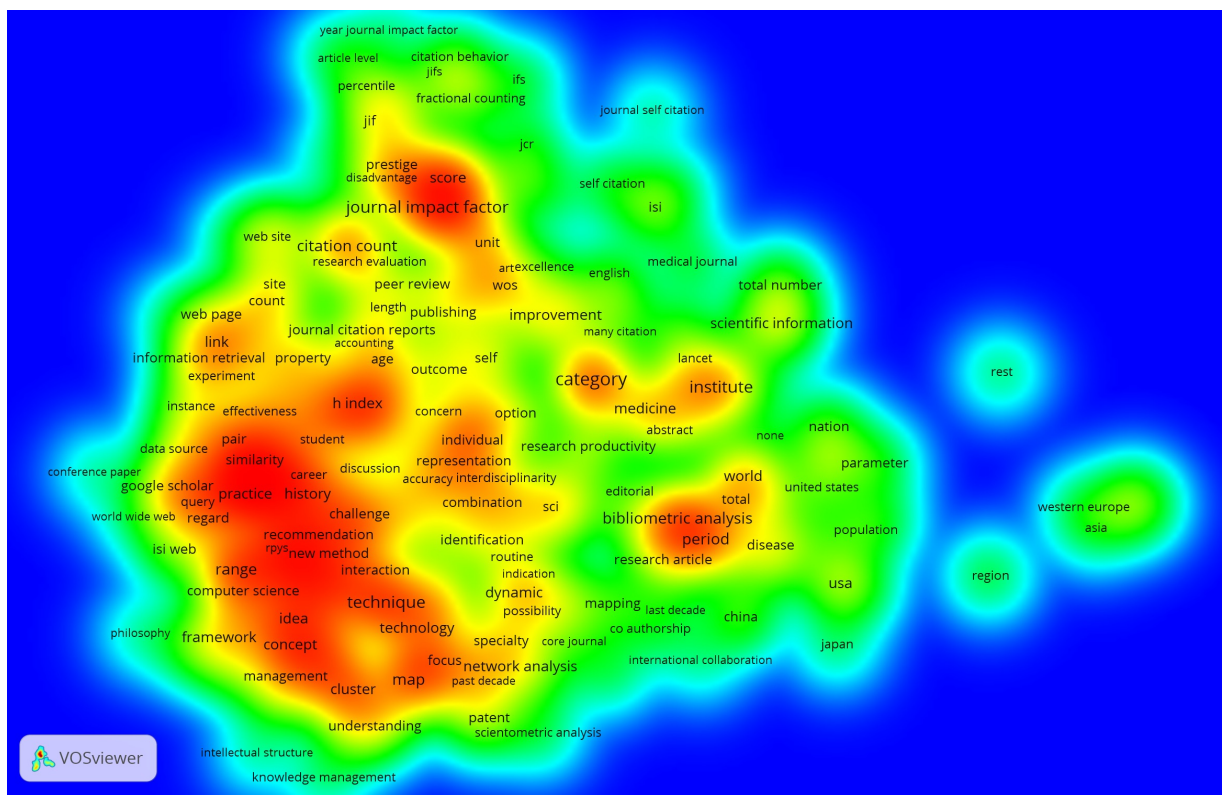

The 18 clusters identified by VOSviewer can be characterized with the following extracted keywords:

(1)     Journal Impact Factor (improvement, citation distribution, percentiles, disadvantages of JIF);

(2) SNIP (Scopus, new indicator, reference list, research field);

(3) h-index (career, age, validity);

(4) PageRank (information retrieval, machine learning, World Wide Web);

(5) Journal Citation Report (JIF score, prestige, Eigenfactor);

(6) Science Citation Index (research article, patent, text mining);

(7) Publication output of regions or nations (Asia, USA, Western Europe);

(8) Rise of output of nations (last decade, China);

(9) Self-citation (individuals, journals, colleagues);

(10) Coverage of databases (books, humanities, citation practices);

(11) Mapping and visualization (technique, clustering, interdisciplinarity, innovation);

(12) Citation networks and prediction or recommendation (biology, chemistry, physics);

(13) Diffusion of ideas (intellectual structure, knowledge domain, scientific collaboration, social network analysis);

(14) Scholarly communication (knowledge management, librarian, practitioner, reader);

(15) Publication and citation analysis in medicine (*Lancet*, *New England Journal of Medicine*, PubMed);

(16) Web links (research productivity and quality, links between universities, search engine);

(17) Digital libraries (future, manuscript, repository);

(18) Peer review and research evaluation (rating, citation frequency, excellence).

The 18 clusters cover a broad range of bibliometric topics, which refer – among other things – to products of providers of bibliometric data (e.g. JCR), indicators (e.g. h-index), units of bibliometric analysis (e.g. nations), types of citations (e.g. self-citations), bibliometric methods (e.g. citation networks), and the use of citations in peer review and research evaluation. The broad range of topics show that EG's papers are cited by publications with a very heterogeneous thematic spectrum.

In the third step of the analysis, we contrast the results on the cited and citing papers with those of the citation context in the citing papers. From the 428 citation contexts that enclose references to EG's publications, a total of 2,347 keywords were extracted of which 184 occurred four or more times. Based on the inclusion criteria, 97 keywords were retained in the map. The following three clusters were identified: 'Journal Impact Factor', 'historiographs', and 'Eugene Garfield as founder of the ISI' (see Figure 3).

Hence, in Figure 3, the same two topics as in Figure 1 can be found (i.e. JIF and historiographs). However, a closer look at these two clusters reveals that more and different keywords are associated with these clusters in Figure 3. For example, the cluster 'JIF' additionally comprises keywords such as assessment, research performance, tenure, normalization, Journal Citation Report (JCR), subject category, and medicine. The differences between the keywords of the JIF cluster in all three networks will be explored in more detail at the end of this section. The cluster 'historiographs' in Figure 3 not only comprises the HistCite software package but also keywords such as idea, structure, citation network, scientific development, research front (a term coined by EG according to Sen, 2017), and DNA (a term reminiscient of EG's work on the history of the DNA, see Small, et al., 2017).

A third cluster emerges in Figure 3 that is not present in Figures 1 and 2: EG as the founder of the ISI. This cluster comprises keywords such as citation indexing, information retrieval, the Social Sciences Citation Index (SSCI), Eugene Garfield, founder, database, and ISI. Hence, this cluster refers to the person EG, whereby similar accomplishments of EG are mentioned as in his obituaries. For example, Braun, Glänzel, and Schubert (2017) praise that EG "founded the Institute for Scientific Information (ISI)" (p. 1) and "created several bibliographic indexes and databases" (p. 1).

Figure 3. Co-occurrence of keywords in citation contexts of papers referencing Eugene Garfield

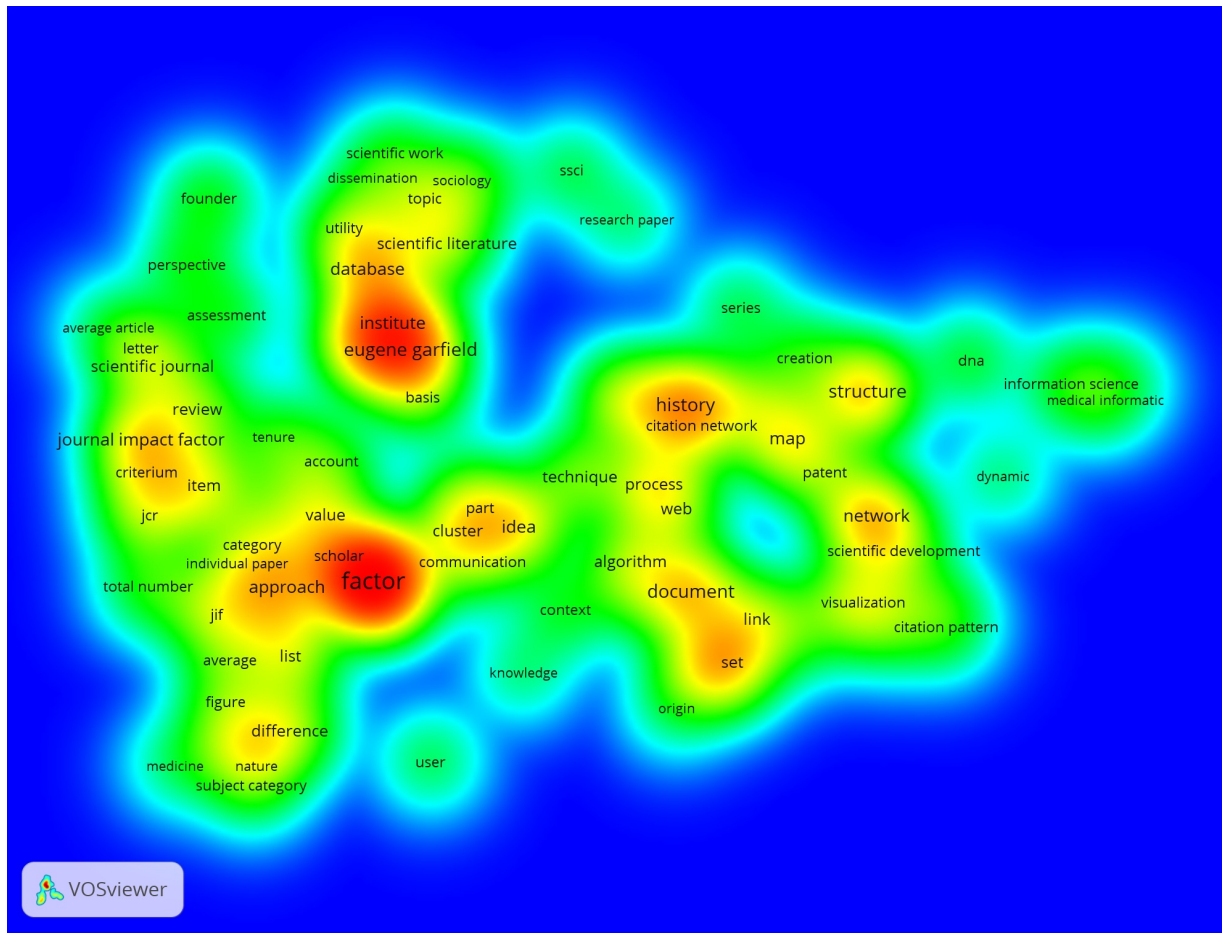

The JIF is a recurring topic and ranks among the most frequent keywords in all three networks. EG "launched the idea of a *journal impact factor,* and created the *Journal Citation Reports* in which citation analysis is applied as a tool for journal evaluation. Again in *Science* he published his most highly cited paper on citation analysis as a tool in journal evaluation" (van Raan & Wouters, 2017).

Thus, we take a closer look at the keywords associated with this cluster and compare the results of the three networks. Table 1 lists the most frequent keywords of the JIF cluster in cited papers, citing papers, and citation contexts. As expected, the lowest number of keywords can be found for the cited papers. With 'journal', 'impact', 'impact factor', and 'journal impact factor', these keywords focus on the indicator itself and do not refer to related topics. However, the keywords from the citing papers and citation contexts contain several links to related topics. For example, the JIF is associated in titles and abstracts of citing papers with keywords such as 'improvement', 'property', 'citation distribution', 'alternative', and 'percentile rank', indicating a relationship between the JIF and the development of alternative indicators.

Table 1. Most frequent keywords (with their number of occurrences in parentheses) of the cluster 'Journal Impact Factor' in cited papers, citing papers, and citation contexts

| **Cited papers** | **Citing papers** | **Citation contexts** |
|---|---|---|
| journal (17) | journal impact factor (29) | factor (60) |
| impact (11) | citation count (21) | journal impact factor (17) |
| impact factor (8) | improvement (17) | approach (14) |
| journal impact factor (6) | normalization (16) | list (13) |
| evolution (5) | statistic (13) | difference (11) |
| number (5) | unit (12) | item (11) |
| use (5) | property (12) | cluster (10) |
| history (4) | significance (11) | review (10) |
|  | citation distribution (9) | period (9) |
|  | weight (9) | scientific journal (9) |
|  | WoS (8) | value (9) |
|  | alternative (8) | normalization (8) |
|  | percentile (7) | account (7) |
|  | percentile rank (7) | category (7) |
|  | IFs (7) | communication (7) |
|  | fractional counting (6) | criterium (7) |
|  | citation behavior (6) | JCR (7) |
|  | selection (6) | JIF (7) |
|  | relevance (6) | total number (7) |

The relationship between the JIF and alternative indicators cannot exactly be found in the keywords from citation contexts. However, in both networks (citing papers and citation contexts) the 'normalization' is among the most frequent keywords. It refers to the problem of the JIF that it cannot be used for cross-field comparisons. Citation data have field-specific patterns which can be normalized by different methods (Bornmann & Marx, 2015).

# 5   Discussion

The turn to new public management and the request for greater efficacy in science has led to the rapid diffusion of quantitative measures of performance and benchmarking (Lamont, 2012). Bibliometrics play an important role in this development. The bibliographic databases initially developed by EG "opened the way to carry out citation analysis of scientific publications on a large scale. Although Garfield was ambivalent about the use of the citation index for research evaluation, the SCI [Science Citation Index] became a powerful source for the measurement of the impact of publications. It enabled the rise of bibliometrics as a policy oriented empirical field" (van Raan & Wouters, 2017). In most of the bibliometric studies published hitherto, the citation is the basic unit of analysis and citations have been simply counted. "A citation from the citing publication to the cited publication counts as a 'vote' for the impact of the paper and aggregate citation statistics are then used to come up with evaluative metrics for measuring scientific impact" (Jha, et al., 2016).

Citation analyses become meaningful if the cited or citing papers are linked with people, ideas, journals, institutions, fields etc. and the data are analyzed with advanced statistical methods (Mingers & Leydesdorff, 2015). The analysis of bibliometric data with network techniques is a good example in this respect, which was initially proposed by Garfield, et al. (1964). In this study, we combine two research lines in bibliometrics: The first line refers to studies in bibliometrics which analyze the context of citations. Many studies in this area have been conducted against the backdrop that it is a sign for a good bibliometric statistic that we receive more than a number as result (Best, 2012). Citation context data can be used to learn more precise details on the citations in a document. For example, the findings of Halevi and Moed (2013) show that "an analysis of citations based on their in- and out-disciplinary orientation within the context of the articles' sections can be an indication of the manner by which cross-disciplinary science works, and reveals the connections between the issues, methods, analysis, and conclusions coming from different research disciplines" (p. 1903). The second line, which we address in this study, concerns the visualization of citation context data as networks. Different methods have been proposed to map various bibliometric data (see an overview in Cobo, López-Herrera, Herrera-Viedma, & Herrera, 2011). In this study, we introduce a new network type which is based on citation context data.

We used the publication set of EG to demonstrate the new network type. According to Wouters (2017), EG's publications "are gold mines for historians" with a very broad spectrum of topics (see an overview in van Raan & Wouters, 2017). We based the study on bibliometric data from MA, since MA provides data on the context of citations. We explored the meaningfulness of results from citation context data and studied whether the citation context is stronger related to the citing or cited document. In other words, we compared the citation context networks with networks which are based on title and abstract words, respectively, of the cited papers (EG's papers) and citing papers. We attempted to identify the meaning of citations to EG's papers by using keyword co-occurrence networks, which are based on extracted words from the citation context. Since this is the first study producing such networks and using citation context data from MA, we focus primarily on opportunities of the approach and secondarily on results.

Our empirical example has demonstrated that this network type leads to meaningful results which characterise how cited studies are perceived. Networks based on titles and abstracts of the citing papers cover a broad range of topics but the overlap with the networks based on the cited papers and based on citation contexts is marginal. First, no cluster and no keyword referring to historiography could be identified. Second, the three clusters representing EG's achievements (i.e. JIF, JCR, SCI) are associated with different keywords in the citation context network than in the other two networks. In particular, the JIF is discussed in relation to other indicators such as SNIP, percentiles, and the h-index, JCR in relation to JIF scores and the Eigenfactor, and the SCI in relation to patents.

Taken as a whole, the comparison of the three networks suggests that papers of EG and citation contexts of paper citing EG are semantically more closely related to each other than to titles and abstracts of papers citing EG. This result accords to our expectation that the citation context network is related to title and abstract keywords from the cited but not citing publications. This is in line with the use of citation data in research evaluation purposes (see section 1). However, we included only a restricted set of EG's papers in this study (see section 3.2). Many papers could not be found in MA. Furthermore, the citation context was not available for many citing papers. Future studies investigating the relationship between cited papers, citing papers, and citation contexts (with co-occurrence networks) should use significantly larger datasets covering broad ranges of subject areas.

## Acknowledgements

We thank the development team of Microsoft Academic for their support and advice.